\documentclass[a4paper,11pt]{article}
\usepackage{amssymb,amsmath,amsthm}
\usepackage{a4wide}
\usepackage{pgf,tikz}
\usetikzlibrary{arrows}
\usepackage{epsfig}
\usepackage{color}
\usepackage{mathdots}

\usepackage{hyperref}

\usepackage{authblk}

\usepackage{subcaption}

\newcommand{\beq}{\begin{equation}}  
\newcommand{\eeq}{\end{equation}}  
\newcommand{\bea}{\begin{eqnarray}} 
\newcommand{\eea}{\end{eqnarray}}   
\newcommand{\bear}{\begin{array}}  
\newcommand{\eear}{\end{array}}

\newtheorem{thm}{Theorem}[section]

\newenvironment{prf}{\trivlist \item [\hskip 
\labelsep {\bf Proof:}]\ignorespaces}{\qed \endtrivlist}

\theoremstyle{definition}

\newtheorem{exa}[thm]{Example}
\newtheorem{remark}[thm]{Remark}

\newcommand{\Z}{{\mathbb Z}}

\newcommand{\R}{{\mathbb R}}

\newcommand{\rd}{\mathrm{d}}

\newcommand\ka{{\kappa}}
\newcommand\al{{\alpha}}
\newcommand\be{{\beta}}
\newcommand\ze{{\zeta}}
\newcommand\gam{{\gamma}}

\newcommand\si{{\sigma}}

\begin{document}


\title{Similarity reductions of peakon equations: the $b$-family} 
\author{Lucy E. Barnes and Andrew N. W. Hone\footnote{ 
e-mail: A.N.W.Hone@kent.ac.uk}\\
School of Mathematics, Statistics \&  Actuarial Science, ~\\ 
University of Kent, ~\\
Canterbury CT2 7FS, U.K.
}

\maketitle

\begin{abstract} 
The $b$-family  is a one-parameter family 
of Hamiltonian partial differential equations of non-evolutionary type, which 
arises in shallow water wave theory. It admits a variety of solutions, 
including the celebrated peakons, which are weak solutions in the form of peaked solitons 
with a discontinuous first derivative at the peaks, as well as other interesting solutions 
that have been obtained in exact form and/or numerically. 
In each of the special cases $b=2,3$ (the Camassa-Holm and Degasperis-Procesi equations, respectively) the equation is completely integrable, in the sense that it admits a 
Lax pair and an infinite hierarchy of commuting local symmetries, but for other values of the parameter $b$ 
it is non-integrable. 
After a  discussion of travelling waves via the use of a reciprocal transformation, which reduces 
to a hodograph transformation at the level of the ordinary  differential equation satisfied by these solutions, 
we apply the same 
technique to the scaling similarity solutions of the $b$-family, and show that when $b=2$ or $3$ this similarity reduction 
is related by a hodograph transformation to particular cases of the Painlev\'e III equation, while for all other choices of $b$ 
the resulting ordinary differential equation is not of Painlev\'e type. 
\end{abstract}

\section{Introduction}

\setcounter{equation}{0}

The one-parameter family of partial differential equations (PDEs) given by 
\beq\label{bfam} 
u_t-u_{xxt} +(b+1)uu_x=bu_xu_{xx}+uu_{xxx}, 
\eeq 
where $b\in\R$ is a parameter, is known as the $b$-family. It was originally introduced in work by one of us with Degasperis 
and Holm \cite{dhh1, dhh2}, in order to analyse the integrable case $b=3$ which had been found a little earlier by 
Degasperis and Procesi \cite{dp}, and facilitate comparison with the celebrated Camassa-Holm case $b=2$, which 
was  derived in the physical context of shallow water theory in \cite{ch1}, 
although its integrability could already be understood within the theoretical 
framework of hereditary symmetries and recursion operators 
described in \cite{ff}. 
It was subsequently shown in \cite{dgh1, dgh2} 
that all of the equations (\ref{bfam}) apart from $b=-1$ are asymptotically equivalent by means of a suitable Kodama transformation, while in \cite{cl} (see also \cite{ivanov}) the equation for parameter values $b\geq 10/11$ or $b\leq -10$ 
was derived from a model of shallow water flowing over a flat bed, with $u$ being the horizontal component 
of fluid velocity at the level line $\theta=\sqrt{\tfrac{11b-10}{12b}}$, $0\leq\theta\leq 1$. 

In the application to shallow water models, the equation (\ref{bfam}) appears with the inclusion of additional 
linear dispersion terms, namely multiples of $u_x$ and $u_{xxx}$. However, such terms can always be removed by 
a combination of a Galilean transformation, going to a moving frame with independent variables $x'=x-vt$, $t'=t$, 
together with a shift to a new dependent variable $u'=u+h$, where the velocity $v$ and background $h$ are constants. 
Thus, for the purposes of what follows, we will work with the dispersionless form of the $b$-family equation, bearing 
in mind that the addition of linear dispersion to (\ref{bfam}) changes the boundary conditions of the solution, so that some of 
the solutions we consider with a non-zero (constant) background correspond to solutions which vanish at infinity when 
dispersion is introduced into (\ref{bfam}). 
In the dispersionless  case, it will be convenient to rewrite (\ref{bfam}) in terms of a momentum density $m$, in the more compact form 
\beq\label{meq} 
m_t +um_x+bu_xm=0, \qquad m=u-u_{xx}. 
\eeq 
In the latter form, the $b$-family can be viewed as a nonlocal evolution equation for $m$, where the 
nonlocality arises from the fact that 
\beq\label{nonloc}
u=g*m=\int_\R g(x-y)m(y)\, \rd y,  
\eeq  
where $g$ is the Green's function for the Helmholtz operator $1-D_x^2$ on the real line, that is 
\beq\label{greens}
g(x) = \frac{1}{2}\exp(-|x|),  
\eeq 
 so that 
$(1-D_x^2)g(x)=\delta(x)$. 

The dispersionless version of the equation (\ref{bfam}), or equivalently (\ref{meq}), is distinguished 
by the remarkable fact, first observed by Camassa and Holm in the case $b=2$,  that with vanishing 
boundary conditions at infinity  it admits weak soliton solutions called peakons, which for any positive 
integer $N$ are given by 
a linear superposition of $N$ peaked solitons, that is   
\beq\label{peakons} 
u(x,t)=\sum_{j=1}^N p_j(t) \exp\big(-|x-q_j(t)|\big), 
\qquad 
m(x,t)=2\sum_{j=1}^N p_j(t) \delta\big(x-q_j(t)\big),
\eeq     
subject to the requirement that the positions $q_j(t)$ and amplitudes $p_j(t)$ satisfy the system of 
ordinary differential equations (ODEs) 
\beq\label{podes}
\frac{\rd q_j}{\rd t}=\frac{\partial \tilde{H}}{\partial p_j} , 
\quad 
\frac{\rd p_j}{\rd t}=-(b-1)\frac{\partial \tilde{H}}{\partial q_j},\qquad j=1,\ldots, N,  
\eeq 
with 
$$ 
\tilde{H}=\frac{1}{2}\sum_{j,k} p_jp_k\exp(-|q_j-q_k|).
$$ 
When $b=2$, the latter ODEs have the form of a Hamiltonian system with $q_j,p_j$ being canonical positions and momenta, 
and $\tilde{H}$ being the Hamiltonian function, and in this particular case Hamilton's equations are also 
completely integrable in the Liouville-Arnold sense. For other values of $b$, it turns out that the ODEs (\ref{podes}) 
can still be considered as a finite-dimensional Hamiltonian system, but with respect to a non-canonical Poisson bracket \cite{hh}, and the two-body problem ($N=2$) is integrable for any $b$. In fact, the ODEs (\ref{podes}) always have 
two first integrals \cite{clp}, given by 
$$H=\sum_{j=1}^N p_j, \qquad P=\prod_{j=1}^N p_j\prod_{k=1}^{N-1}(1-e^{-|q_k-q_{k+1}|})^{b-1},
$$ 
but it 
seems almost certain that the equations of motion for the peakons with $N>2$ can only be explicitly integrated in the special 
cases $b=2$ and $b=3$, since the exact solutions obtained in   \cite{bss1, bss2} and \cite{ls1, ls2}, respectively, rely heavily 
on the use of an appropriate spectral problem derived from the underlying Lax pair for the corresponding integrable PDE 
in each of those cases.   

For any $b\neq 0,1$ the equation (\ref{meq}) can be derived from the least action principle $\delta S=0$ with 
$S=\int\int\mathcal{L}\,\rd x\,\rd t$, where the Lagrangian density is 
\beq\label{lag} 
\mathcal{L} =\frac{\varphi_t}{2\varphi_x}\Big((\log\varphi_x)_{xx}+1\Big)-\frac{\varphi_x^b}{b-1}, 
\eeq  
which arises by rewriting the equation as the conservation law 
\beq\label{cons}
p_t+(up)_x=0, \qquad p=m^{1/b},  
\eeq  
and then introducing $\varphi$ as a potential such that 
\beq\label{pot}
p=\varphi_x, \qquad u=-\frac{\varphi_t}{\varphi_x}.  
\eeq 
An appropriate Legendre transformation leads to the Hamiltonian form of the equation (\ref{meq}), namely 
\beq\label{ham}
m_t = \frac{1}{b-1} \, B\, \frac{\delta H}{\delta m},  
\eeq 
with 
\beq\label{hamn}
H=\int m\, \rd x, \qquad B=(bmD_x+m_x)(D_x-D_x^3)^{-1}(-bD_xm+m_x),  
\eeq 
valid for any $b\neq 1$; for two different proofs of the Jacobi identity for the skew-symmetric operator $B$, 
see \cite{hh, hw}. (The same Hamiltonian operator $B$ works for $b=1$ in (\ref{ham}) with the 
replacement $(b-1)^{-1} H \to 
\int m \log m\,\rd x$.) This operator has two independent Casimir functionals, namely 
\beq\label{cas2}
C_1=\int m^{1/b}\, \rd x, \qquad
C_2= \int m^{-1/b} \left( \frac{m_x^2}{b^2m^2}+1\right)\, \rd x, 
\eeq   
where the density of the first one  corresponds to the conservation law (\ref{cons}). 
The matter of determining appropriate classes of solutions for which these functionals are well-defined, or require 
appropriate regularization, and how this depends on the value of $b$, is a delicate one. (See \cite{hl} for instance, 
where a Banach subspace of a weighted Sobolev space was considered 
in order to prove an orbital stability property of stationary solutions when $b<-1$.)

The $b$-family has various interesting geometric properties, in addition to its Lagrangian and Hamiltonian structure. 
There is the conservation equation 
\beq \label{conseq}
m(q,t) q_x^b=m(x,0),  
\eeq  
where $x\mapsto q(x,t)$ is a diffeomorphism of $\R$ defined by the initial value problem 
$$ 
q_t=u(q,t), \qquad q(x,0)=x.
$$ 
The equation (\ref{conseq}) holds for all $t$ in the domain of existence of the solution of (\ref{meq}), and 
for solutions with $m(x,0)>0$ it implies that $m$ remains positive as long as the solution exists. Moreover, if the periodic solutions of the equation 
are considered, taking $S^1$ instead of the real line $\R$, then the $b$-family equation can be regarded as the 
geodesic equation  for a suitable connection on the diffeomorphism group of the circle \cite{ek} (for generic $b$, this is a non-metric connection, but the case $b=2$ gives an Euler-Poincar\'e equation for geodesics with respect to the $H^1$ metric 
\cite{m}).  

The case of solutions with positive $m>0$ (or with fixed sign everywhere) is especially relevant in what follows, as 
it allows the definition of the reciprocal transformation 
\beq\label{recip} 
\rd X=p\,\rd x -up\,\rd t, \qquad \rd T=\rd t
\eeq 
associated with the conservation law (\ref{cons}), which transforms (\ref{meq}) to a PDE of third order for $p=p(X,T)$ 
as a function of the  independent variables $X,T$, namely 
\beq\label{pXT} 
\frac{\partial }{\partial T}\left(\frac{1}{p}\right)+\frac{\partial }{\partial X}\Big(p(\log p)_{XT} -p^b\Big)=0
\eeq 
(where, by an abuse of notation, we are using the same letter $p$ for the dependent variable, considered as a function of the new variables $X,T$). The equation (\ref{pXT}) can naturally be regarded as an extension of the sine-Gordon equation 
when $b=2$, or of the Tzitzeica equation when $b=3$ (up to replacing $m=p^b\to -p^b$, details can be found in 
\cite{hw}), but for other values of $b$ it fails the Painlev\'e test 
\cite{h}, which is consistent with the results of other integrability tests applied to the $b$-family \cite{dp, mn}.  

Any reciprocal transformation sends a conservation law in the original independent variables to another conservation law in terms of the new variables. 
Hence  
there is another way to rewrite the equation (\ref{pXT}) in conservation form, which corresponds to applying the reciprocal transformation (\ref{recip}) to 
the conservation law associated with the conserved density for the Casimir $C_2$ in (\ref{cas2}): for any $b\neq 1$, we have 
\beq\label{VT}
\frac{\partial V}{\partial T}+\frac{b}{2(b-1)} \frac{\partial }{\partial X}\big(p^{b-1}\big)=0, 
\eeq    
where the quantity $V$ is defined in terms of $p$ by 
\beq\label{ep}
pp_{XX}-\tfrac{1}{2}p_X^2 +2Vp^2+\tfrac{1}{2}=0.
\eeq 
Regarded as an ODE for $p$ as a function of $X$, which $V$ given, the latter equation is known as Ermakov's equation \cite{ermakov}, or the Ermakov-Pinney equation 
(see \cite{hw} for further references). Then $V$ (up to a sign) can be interpreted as the potential for  a Schr\"odinger operator, and the general solution of 
(\ref{ep}) can be written in the form 
of a product 
\beq\label{psol} 
p=\psi_+\psi_-, \qquad \mathrm{for} \quad (D_X^2+V)\psi_{\pm}=0, \qquad \mathrm{with} \quad W(\psi_+,\psi_-)^2=1, 
\eeq 
where 
$$ 
W (\psi_+,\psi_-)=\psi_{+,X}\psi_- -\psi_{-,X}\psi_+ 
$$ 
is the Wronskian of the two solutions of the Schr\"odinger equation. 
The logarithmic derivatives of these two wave functions can be written in terms of $p$, as 
$$ 
\frac{\rd}{\rd X}\log\psi_{\pm}=\frac{p_X\pm 1}{2p}.
$$
We shall make use of the representation (\ref{psol}) of $p$ when we consider similarity reductions of (\ref{pXT}) in the sequel.   

Some time ago, Holm and Staley did a series of extensive numerical studies of the solutions of the $b$-family, and observed remarkable 
bifurcation phenomena controlled by the parameter $b$ \cite{hs1, hs2}: given initial data vanishing at infinity, they found a train of peakons was produced for 
$b>1$; but for $-1<b<1$ the same initial value problem appears to form something that resembles the ramp/cliff profile seen in Burgers' equation, given by the similarity solution 
(ramp) 
\beq\label{ramp} 
u(x,t)=\frac{x}{(b+1)t} 
\eeq 
in a compact region, joined to a rapidly decaying cliff; while for $b<-1$ the initial profile develops into a train of lefton solutions, consisting of solitary waves that move to the left 
before becoming stationary (see (\ref{lefton}) below for the explicit  form of a lefton solution). As such, the $b$-family (at least for $|b|>1$, in the range where peakons/leftons 
appear) seems to provide support 
for the soliton resolution conjecture (see e.g. \cite{tao}),   which says that for any suitable dispersive evolutionary PDE (not necessarily integrable), generic initial data should decompose into 
a train of solitary waves together with radiation which decays to zero as $t\to \infty$. Apart from intensive studies of the 
integrable cases $b=2,3$, further analytical and numerical support for the behaviour reported by Holm and Staley has taken a while 
to materialize: an orbital stability result for a single lefton when $b<-1$ was proved in \cite{hl}, while there are various well-posedness and rigidity results (see \cite{inci, molinet}
and references);
yet linear stability/instability results for peakons, ramp/cliff solutions and leftons across
the full range of $b$ values have been found only very recently \cite{cpkl, lp}.

In this paper, we are concerned with describing scaling similarity solutions of the $b$-family (\ref{bfam}) which generalize the ramp (\ref{ramp}). 
Our main result is that (for $b\neq 0$) such solutions satisfy an autonomous ODE of  third order, which is related via a hodograph transformation to a 
non-autonomous second order ODE that closely resembles 
the third Painlev\'e equation 
\beq\label{piii}
\frac{\rd^2 w}{\rd \ze^2}=\frac{1}{w}\left(\frac{\rd w}{\rd \ze}\right)^2-\frac{1}{\ze}\left(\frac{\rd w}{\rd \ze}\right)+\frac{1}{\ze}\big( \al w^2+\be) +\gam w^3 +\frac{\delta}{w}.
\eeq
Unlike the latter, %
for generic values of $b$ the second order equation we find does not have the Painlev\'e property, with the exception of the special values $b=2,3$, which turn out to correspond to 
particular instances of  (\ref{piii}). (Some details for the Camassa-Holm case $b=2$ were already derived in \cite{hach}.) 

Our method for deriving the hodograph transformation for   scaling similarity solutions is based on reduction of 
the reciprocal transformation (\ref{recip}), so as a warm-up exercise, in the next section we show how the same method 
works in the slightly more straightforward context of the smooth travelling wave solutions of (\ref{bfam}), which can be 
reduced to a quadrature.      After applying the hodograph transformation, in the cases $b=2,3$ the travelling waves 
are given explicitly in parametric form in terms of Weierstrass functions, for which we give full details (omitting a 
discussion of the degenerate case of vanishing discriminant, $g_2^3-27g_3^2=0$, which 
produces the smooth 1-soliton solution of Camassa-Holm/Degasperis-Procesi, described elsewhere \cite{ch2, hach, matsuno}).
We also present the periodic travelling waves for $b=-1$, which are given parametrically in terms of trigonometric functions. 
The third section is devoted to the scaling similarity solutions of (\ref{bfam}), and the corresponding 
parametric formulae obtained via a    hodograph transformation. Once again, after describing the general case, we 
focus on the special parameter values $b=2,3$ and clarify the connection with particular cases of the third Painlev\'e equation, 
before ending with a brief section of conclusions.

\section{Travelling waves and hodograph transformation}

\setcounter{equation}{0}

We start by considering travelling waves of (\ref{bfam}), setting 
$$ 
u(x,t)=U(z), \qquad m(x,t)=M(z), \qquad z=x-ct, 
$$ 
where $c$ is the wave velocity, and we will also write $P(z)$ for the quantity $M^{1/b}$. The conservation 
law (\ref{cons}) becomes a total $z$ derivative, so integrating this we obtain 
\beq\label{up}
(U-c)P+d=0,  
\eeq
where $d$ is an integration constant. 
Henceforth we will assume that $d\neq 0$, as the case $d=0$ implies that either $U=c$, a constant, or $P=0$, if we are considering 
smooth solutions; but the 1-peakon solution with $U=c\exp (-|z-z_0|)$, $M=2c\delta(z-z_0)$ ($z_0$ arbitrary) can 
be viewed as a weak limit of strong (analytic) solutions with $d=0$ (see the discussion in \cite{ch2} or \cite{lo}, 
for instance).

Then from the formula relating $m$ and $u$, as in (\ref{meq}), we find 
\beq\label{upb}
U_{zz}-U+P^b=0. 
\eeq 
If we substitute $U=c-d/P$ from (\ref{up}) into the latter, then we obtain an equation of second order for $P$, 
namely 
\beq\label{trav1}
P_{zz}-2\frac{P_z^2}{P}-cd^{-1}P^2+P +d^{-1}P^{b+2}=0. 
\eeq 
This equation can be integrated to yield an equation of first order, that is 
\beq\label{trav1st}
P_z^2=F(P), \qquad \mathrm{where} \qquad F(P)= P^2-2cd^{-1}P^3+eP^4 - \frac{2P^{b+3}}{d(b-1)}, 
\eeq  
with $e$ being another integration constant (and it is necessary to assume $b\neq 1$, otherwise a term with $\log P$ appears). 
Thus the determination of travelling waves reduces to the quadrature $\int \rd P /\sqrt{F(P)}=z+\,$const.  

Observe that the assumption $d\neq 0$ puts constraints on the boundary conditions of the travelling waves, depending on the  sign of $b$. When $b>0$, the combination of  the relation (\ref{upb}) with $U=c-d/P$ implies that there is no smooth 
solution $U(z)$ that vanishes at infinity, so that only periodic or unbounded waves are possible in that case. 
(This can also be seen by 
considering the phase portrait of (\ref{trav1}) in the $(P,P_z)$ plane.) However, when $b<0$ it is possible to have solutions 
with $U\to 0$ and $P\to \infty$ as $|z|\to\infty$: for instance, when $b<-1$ and $c=0$ there are the stationary lefton solutions, given explicitly by 
\beq\label{lefton}
U(z)=A\Big(\cosh\gam(z-z_0)\Big)^{-\frac{1}{\gam}}, 
\, 
M(z)=\frac{A(1-b)}{2} \Big(\cosh\gam(z-z_0)\Big)^{\frac{b}{\gam}}, \,
\gam=-\frac{b+1}{2}, 
\eeq  
where $A$ is an arbitrary constant that can be written in terms of $d$ and $b$, corresponding to setting $c=0$ and $e=0$ 
in (\ref{trav1st}).

It is instructive to see how the same solutions arise via reduction of the Lagrangian density (\ref{lag}) for the PDE. If 
we replace $p=\varphi_x \to \phi_z=P$ and then note that we require $u=-\varphi_t/\varphi_x\to U=c-d/P=c-d/\phi_z$, 
then we obtain the Lagrangian 
$$ 
L = -\frac{1}{2}\left( c-\frac{d}{\phi_z}\right)\, \Big((\log\phi_z)_{zz}+1\Big)-\frac{\phi_z^b}{b-1}, 
$$ 
but we can subtract off the terms involving $c$ (a constant plus a total $z$ derivative) since these will not affect 
the Euler-Lagrange equation, to obtain $L\to\bar{L}$, 
where 
\beq\label{redL} \bar{L}=
\frac{d}{2\phi_z}\Big((\log\phi_z)_{zz}+1\Big)-\frac{\phi_z^b}{b-1}. 
\eeq  
Because $\phi$ does not appear, the Euler-Lagrange equation for (\ref{redL}) is a total derivative: 
$$ 
\frac{\rd}{\rd z}\left(-\frac{\rd^2}{\rd z^2}\Big(\frac{\partial \bar{L}}{\partial \phi_{zzz}}\Big)+
\frac{\rd}{\rd z}\Big(\frac{\partial \bar{L}}{\partial \phi_{zz}}\Big)-
\frac{\partial \bar{L}}{\partial \phi_{z}}\right) = 0 .
$$
The above equation can be integrated, with an integration constant $C$, or we can replace 
$\phi_z$ and its derivatives in terms of $P$ and add on a term with $C$ as a  Lagrange multiplier: 
$\bar{L}(\phi_z,\phi_{zz},\phi_{zzz})\to \bar{L}(P,P_z,P_{zz})+C(P-\phi_z)$; then the 
leading terms involving derivatives are $\tfrac{d}{2}(P_{zz}P^{-2}-P_z^2 P^{-3})=
\tfrac{d}{2}\big((P_zP^{-2})_z+P_z^2P^{-3}\big)$, so removing the total derivative leads to an equivalent 
Lagrangian in terms of $P$, that is 
\beq\label{Plag}
\hat{L}=\frac{d}{2} \left(\frac{P_z^2}{P^3} +\frac{1}{P}\right) -\frac{P^b}{b-1}+CP.
\eeq 
The Euler-Lagrange equation for $P$ obtained from $\hat{L}$ is different from (\ref{trav1}), 
but applying a Legendre transformation to (\ref{Plag}) yields the conjugate momentum to $P$ 
and a conserved Hamiltonian, 
namely 
$$ 
\pi  =\frac{\partial\hat{L}}{\partial P_z}=\frac{dP_z}{P^3}\implies h=\pi P_z-\hat{L}
$$ 
so that on a fixed level set $h=\,$const we have 
$$ 
h=\frac{d}{2}\left(\frac{P_z^2}{P^3}-\frac{1}{P}\right)+\frac{P^b}{b-1}-CP, 
$$ 
and this agrees with  (\ref{trav1st}) when we identify $c=-h$, $e=2C/d$. 

By considering how the travelling wave solutions of (\ref{meq}) behave under the action 
of the reciprocal transformation (\ref{recip}), it is not hard to see that the roles of the parameters $c,d$ are reversed: 
for travelling waves of (\ref{meq}) with velocity $c$, the parameter $d$ appears as an integration constant from (\ref{cons}), while 
 travelling waves of (\ref{pXT}) with velocity $d$ satisfy the same relation $U=c-d/P$, but now $c$ appears as an integration constant by 
rewriting (\ref{pXT}) as $(p^{-1})_T-u_X=0$ and setting 
$$p(X,T)=P(Z), \qquad u(X,T)=U(Z), \qquad Z=X-dT,$$ 
then integrating the reduced equation $-\tfrac{\rd}{\rd Z}(U+dP^{-1})=0$. Hence the reciprocal transformation 
reduces to a hodograph transformation, that is 
\beq\label{hod}
\rd Z=\rd (X-dT)=p\,\rd x -(up+d)\,\rd t=P(z) \rd (x-ct)=P(z)\,\rd z, 
\eeq where  note that once again we will abuse notation by using the same letter $P$ for the dependent 
variable viewed as a function of either argument ($z$ or $Z$), i.e.\ $P(Z)\equiv P\big(z(Z)\big)$.    
Hence, under this hodograph transformation, the derivatives transform as 
$\tfrac{\rd}{\rd z}=P\tfrac{\rd}{\rd Z}$, so that from (\ref{trav1st}) the first order ODE for $P(Z)$ is just 
\beq\label{hod1st}
\left(\frac{\rd P}{\rd Z}\right)^2=F^*(P), \qquad \mathrm{where} \qquad F^*(P)= 1-2cd^{-1}P+eP^2 - \frac{2P^{b+1}}{d(b-1)}.
\eeq  

One can also obtain an ODE of second order for $P(Z)$ by starting from the action $\bar{S}=\int \bar{L} \,\rd z$,  after replacing $\bar{L}\to\bar{L}+\,$const and 
$\rd z\to P^{-1}\rd Z$ from (\ref{hod}), to obtain a new action $S^*=\int L^*\,\rd Z$, and then (\ref{hod}) is satisfied on each level set of an appropriate 
Hamiltonian, obtained via a Legendre transformation applied to $L^*$. Also, the $X$ derivatives in the Ermakov equation (\ref{ep}) all become $Z$ derivatives under this reduction, so 
that we may write the potential  $V=V(Z)$ as  
\beq\label{Veq} 
V=-\frac{1}{2P}\left(\frac{\rd^2 P}{\rd Z^2}\right)+\frac{1}{4P^2}\left(\Big(\frac{\rd P}{\rd Z}\Big)^2-1\right);
\eeq 
but then for the travelling wave solutions, the conservation law (\ref{VT}) becomes a total $Z$ derivative, and this can be integrated to yield 
\beq\label{Vform}
V=\frac{b}{2d(b-1)}P^{b-1} -\frac{e}{4}, 
\eeq  
where value of the integration constant is found by comparing (\ref{Veq}) with (\ref{hod1st}). Hence we arrive at the 
main result of this section.

\begin{thm}\label{travhod} 
The travelling wave solutions $u(x,t)=U(z)$, $z=x-ct$ of the $b$-family equation (\ref{meq}) for $b\neq 0,1$, with constant $d\neq 0$, 
are given in parametric form by $U(Z)=c+d/P(Z)$, $z=z(Z)$, where 
\beq\label{params}
\int^{P(Z)} \frac{\rd s}{\sqrt{F^*(s)}}=Z+\mathrm{const}, 
\qquad 
z(Z)=\log\left(\frac{\psi_+(Z)}{\psi_-(Z)}\right)+\mathrm{const}, 
\eeq
with  $F^*$ defined by (\ref{hod1st}), 
where $\psi_{\pm}$ are two independent solutions of the same Schr\"odinger equation, 
\beq \label{schro}
\left(
\frac{\rd^2}{\rd Z^2}+V(Z) 
\right) \psi_\pm =0 
\eeq 
having Wronskian $W(\psi_+,\psi_-)=1$, subject to the requirement that 
$P(Z)=\psi_+(Z)\psi_-(Z)$, 
and $V(Z) $ is given in terms of $P=P(Z)$ by (\ref{Vform}). 
\end{thm}

\begin{prf}
The quadrature for $P(Z)$ in (\ref{params}) follows immediately from (\ref{hod1st}). To obtain the formula for $z(Z)$, first of all note that 
from the theory of Ermakov's equation (\ref{ep}), if $V(Z)$ is fixed by (\ref{Vform}) then the relation (\ref{Veq}) implies that there is a pair of 
independent solutions of (\ref{schro}) with Wronskian 1 such that 
$P=\psi_+\psi_-$. Then we have 
$$ 
\rd z=\frac{1}{P(Z)}\,\rd Z=\frac{W(\psi_+, \psi_-)}{\psi_+\psi_-}\, \rd Z=\rd \log \left(\frac{\psi_+(Z)}{\psi_-(Z)}\right), 
$$ 
and the result follows.
\end{prf}

\begin{exa} {\bf The Camassa-Holm equation:} In the case $b=2$, for the analytic travelling wave solutions of the Camassa-Holm equation, 
the hodograph-transformed ODE is 
$$
\left(\frac{\rd P}{\rd Z}\right)^2 =1-2cd^{-1}P+eP^2-2d^{-1}P^3
$$ 
which is solved in terms of elliptic functions. Up to the freedom to replace $Z\to Z+\,$const (which 
is useful to exploit, shifting by a suitable half-period in order to obtain non-singular 
solutions that are periodic and bounded for real $Z$), the solution can be written 
in terms of the Weierstrass function $\wp(Z)=\wp(Z;g_2,g_3)$ with arbitrary invariants $g_2,g_3$ and 
another arbitrary parameter $W$, as 
\beq\label{Pb2} 
P(Z)=\frac{\wp(Z)-\wp(W)}{\wp'(W)},
\eeq    
with the coefficients in the ODE for $P(Z)$ being given by 
$$ 
c=\frac{\wp''(W)}{2\wp'(W)^2}, \quad d=-\frac{1}{2\wp'(W)}, 
\quad e=12\wp(W), 
$$ 
and the equation (\ref{Vform}) gives 
$$ 
V=d^{-1}P-\frac{e}{4}=-2\wp(Z)-\wp(W),
$$ 
so that the Schr\"odinger equation (\ref{schro}) corresponds to the simplest case of Lam\'e's equation, 
and the two independent solutions with Wronskian 1 are given in terms of the Weierstrass sigma function by 
$$ 
\psi_\pm (Z)=\frac{1}{\sqrt{\wp'(W)}}\, \frac{\si(W\mp Z)}{\si(W)\si(Z)}\,\exp\big(\pm \zeta(W)\, Z\big),
$$ 
and these satisfy $P=\psi_+\psi_-$. Thus, up to shifting by an arbitrary constant,  the travelling wave variable $z$ for the original equation 
has the form 
$$ 
z(Z) =\log \left(\frac{\si (W-Z)}{\si(W+Z)}\right)+2\zeta(W)\, Z.
$$ 
This explicit parametric form for the periodic travelling waves of Camassa-Holm was 
given in \cite{hach}. For higher genus analogues, corresponding to finite-gap solutions 
of Camassa-Holm, see \cite{cm}, for instance.
\end{exa}

\begin{exa}  {\bf The Degasperis-Procesi equation:} In the case $b=3$, for the analytic travelling wave solutions of the Degasperis-Procesi equation, 
the hodograph-transformed ODE is 
\beq\label{b3hod} 
\left(\frac{\rd P}{\rd Z}\right)^2 =1-2cd^{-1}P+eP^2-d^{-1}P^4
\eeq 
which defines a curve of genus one in the $(P,P_Z)$ plane, and is solved in terms of elliptic functions. By making a 
birational transformation from the quartic curve defined by (\ref{b3hod}) to a Weierstrass cubic, we find that the solution is given 
explicitly by 
\beq\label{b3sol} 
P(Z)=\frac{1}{\al\wp'(W_1)} \, \left(\frac{\wp(Z)-\wp(W_1)}{\wp(Z)-\wp(W_2)}\right),  
\eeq 
being specified by the three parameters $g_2,g_3,W_2$, where $\al$ and the quantity $W_1$ that fixes the zeros of $P$ are 
determined  by 
\beq\label{alform}
\al = -\tfrac{1}{2}\, \frac{\wp''(W_2)}{\wp'(W_2)^2}=\big(\wp(W_1)-\wp(W_2)\big)^{-1}, 
\eeq 
while the coefficients in (\ref{b3hod}) are fixed  by 
\beq\label{de}
e=12\wp(W_2)-\tfrac{3}{2}\, \left(\frac{\wp''(W_2)}{\wp'(W_2)}\right)^2, 
\qquad
d=-\frac{16\wp'(W_2)^6}{\wp'(W_1)^2\wp''(W_2)^4}, 
\eeq 
$$
\frac{c}{d}=\frac{8}{\wp'(W_1)}\left(\frac{\wp'(W_2)^4}{\wp''(W_2)^2} 
-3\frac{\wp(W_2)\wp'(W_2)^2}{\wp''(W_2)}+\tfrac{1}{4} \wp''(W_2)\right). 
$$
To obtain the original travelling wave variable $z$ for Degasperis-Procesi parametrically in terms of $Z$, note that 
we may write 
$$ 
P^{-1} =\al \wp'(W_1)+ \frac{\wp'(W_1)}{\wp(Z)-\wp(W_1)}, 
$$ 
and then by standard elliptic function identities this can be integrated with respect to $Z$ to yield 
\beq\label{b3parz}
z(Z) = \big( \al \wp'(W_1)+2\zeta(W_1)\big)Z + \log\left( \frac{\si (W_1-Z)}{\si (W_1+Z)}\right) 
\eeq   
(up to a constant). 
Having obtained the explicit form of $z(Z)$, we can then apply Theorem \ref{travhod} in reverse, 
writing $z(Z)=\log(\psi_+/\psi_-)$, $P=\psi_+\psi_-$ to find that 
\small
\beq\label{psib3}
\psi_\pm (Z)= \frac{1}{\sqrt{\al \wp'(W_1)}}
\, \frac{\si (W_1 \mp Z)}{\si(W_1)\si(Z)}\, \big(\wp(Z)-\wp(W_2)\big)^{-1/2} \,\exp\Big(\pm \big(\zeta(W_1)+\tfrac{1}{2}\al\wp'(W_1)\big)\, Z\Big)
\eeq 
\normalsize
have Wronskian 1 and satisfy the same linear equation 
$$
\left(
\frac{\rd^2}{\rd Z^2}+V(Z) 
\right) \psi_\pm =0 , \qquad V=\frac{3}{4d}\, P^2 -\frac{e}{4},  
$$
from (\ref{Vform}), with the potential given explicitly by 
\beq\label{Vb3} 
V(Z)=\frac{3}{4d\al^2 \wp'(W_1)^2} \, \left(\frac{\wp(Z)-\wp(W_1)}{\wp(Z)-\wp(W_2)}\right)^2-\frac{e}{4}.
\eeq 
The linear equation for $\psi_\pm$ as given by (\ref{psib3}) can be verified directly by rewriting 
it as 
$$
V=-\frac{\rd^2}{\rd Z^2}\log \psi_\pm -\left(\frac{\rd}{\rd Z}\log \psi_\pm\right)^2, 
$$ 
and then noting that both the left-hand and right-hand sides above are 
elliptic functions 
of $Z$ with double poles at points congruent to $\pm W_2$ modulo the 
period lattice of the Weierstrass curve, and nowhere else, with the same leading order
Laurent expansions 
$$-\tfrac{3}{4} (Z\mp W_2)^{-2}+O(1) \quad  \mathrm{as} \, \, Z\to \pm W_2,$$ 
and comparing the value of the function on each side at $Z=0$ we find the identity 
$$ 
\frac{3}{4d\al^2 \wp'(W_1)^2}-\frac{e}{4}=\frac{1}{\al}-\frac{1}{4}\al^2  \wp'(W_1)^2, 
$$ 
which is a consequence of (\ref{de}) and the given expression (\ref{alform}) for $\al$ in terms of elliptic functions 
with argument $W_2$. 
\end{exa}

\begin{exa} {\bf Genus zero solutions for $b=-1$:} In the case $b=-1$, the equation (\ref{hod1st}) becomes 
\beq\label{bm1} 
\left(\frac{\rd P}{\rd Z}\right)^2 =1+d^{-1}-2cd^{-1}P+eP^2, 
\eeq 
which defines a curve of genus zero (a conic) in the $(P,P_Z)$ phase plane. 
If we rule out the case of parabolae ($e=0$) then there are two types of solution for $P(Z)$: unbounded solutions given in terms of hyperbolic functions, when the curve is a hyperbola, and bounded 
periodic solutions, when the curve is an ellipse. We focus on the latter, and consider solutions of the form 
\beq\label{sinsol}
P(Z)=A+B \sin (\gam Z), 
\eeq  
with parameters $A\geq|B|>0$, $\gam>0$, so that $P\geq 0$. This corresponds to taking parameters 
$$
e=-\gam^2, \quad d=-\frac{1}{1+\gam^2(A^2-B^2)}, \quad c=Ade
$$
in (\ref{bm1}). Upon integrating $P^{-1}$ with respect to $Z$, the travelling wave variable for (\ref{meq}) is found to be 
$$ 
z(Z)=\frac{2}{\gam \sqrt{A^2-B^2}} \mathrm{arctan}\left(\frac{A\tan(\tfrac{1}{2}\gam Z)+B}{\sqrt{A^2-B^2}}\right), 
$$ 
up to a constant. By Theorem \ref{travhod} the latter can be rewritten as a logarithm of the ratio of two independent solutions of 
(\ref{schro}) with potential 
$$ 
V(Z) =- \frac{1+\gam^2(A^2-B^2)}{4\big(A+B\sin(\gam Z)\big)^2}+\frac{\gam^2}{4}, 
$$
but we omit further details. 
\end{exa}

\section{Scaling similarity reductions and Painlev\'e equations}

\setcounter{equation}{0}

Each member of the $b$-family of equations (\ref{bfam}), apart from the case $b=0$, admits a scaling
similarity reduction, which is obtained by taking 
\beq\label{ured} 
u(x,t)=t^{-1}U(z), \qquad z=x+ab^{-1}\log t,
\eeq 
where $a$ is an arbitrary parameter. For the variable (momentum density) $m$ in (\ref{meq}), this means that 
we may write 
\beq\label{mpred}
m(x,t)=t^{-1}M(z), \quad M=U-U_{zz} \implies p=m^{1/b}=t^{-1/b}P(z).
\eeq 
Under this reduction, the equation in its original form (\ref{bfam}) reduces to an autonomous ODE of third order for $U(z)$,
namely 
\beq\label{third} 
(U+ab^{-1})(U_{zzz}-U_z)+(bU_z-1)(U_{zz}-U)=0.
\eeq 
The ramp profile (\ref{ramp}) for $b\neq -1$ corresponds to the solution 
\beq\label{Uramp} 
U(z)=\frac{z-z_0}{b+1}
\eeq 
when $a=0$ (with $z_0$ being an arbitrary choice of origin for the ramp).  
As it stands, in general there appears to be no way to integrate the equation (\ref{third}) further. However, by exploiting the reciprocal transformation 
(\ref{recip}), it is possible to obtain the solutions of this equation 
in parametric form from the solutions 
of a non-autonomous ODE of second order that is related via a hodograph transformation. 

\begin{remark}
If we replace $ab^{-1}\to\bar{a}$ with $\bar{a}$ arbitrary, then the reduction (\ref{ured}) 
still makes sense for $b=0$. In that case, $U(z)$ satisfies an ODE of the same form as (\ref{third}), 
i.e.\ 
$$ 
(U+\bar{a})(U_{zzz}-U_z)-U_{zz}+U=0. 
$$
However, our subsequent analysis no longer makes sense for $b=0$, because it is based on the 
dependent variable $P=M^{1/b}$, as in (\ref{mpred}). 
\end{remark} 

The key observation is that, for $b\neq 0$, the PDE (\ref{pXT}) admits the similarity reduction 
\beq\label{pred} 
p(X,T)=T^{-1/b}P(Z), \qquad Z=XT^{1/b}, 
\eeq 
and if the PDE is written in conservation form as $(p^{-1})_T=u_X$, then under this reduction we find 
$u(X,T)=p^b-p(\log p)_{XT}=T^{-1}U(Z)$, where 
$U=P^b-b^{-1}P\big(Z(\log P)_{ZZ}+(\log P)_Z\big)$.
Thus, after removing a factor of $T^{-1}$, the reduced equation becomes a total $Z$ derivative, 
that is 
$$ 
\frac{\rd}{\rd Z}\big(b^{-1}ZP^{-1}-U\big)=0, 
$$ 
which integrates to yield 
\beq\label{upaform}
U(Z)=\frac{1}{b}\left(\frac{Z}{P(Z)}-a\right), 
\eeq
with $a$ being an arbitrary integration constant.  
Upon replacing $U$ in terms of $P$ and its $Z$ derivatives, this gives the second order equation 
\beq\label{logPZ}
P\frac{\rd}{\rd Z} \Big(Z (\log P)_Z\Big) - bP^b+\frac{Z}{P}-a=0.  
\eeq 
At the level of the reciprocal transformation (\ref{recip}), this gives a relation between the similarity reductions of (\ref{bfam}) and (\ref{pXT}): 
we can identify the parameter $a$, which plays a different role in these two reductions, to find that the similarity variable $z$ in (\ref{ured}) satisfies
\begin{align*}%
\rd z  &=\rd (x+ab^{-1}\log t) =p^{-1}\rd X +u\,\rd T +ab^{-1}T^{-1}\rd T\\
&=P(Z)^{-1}T^{1/b}\rd X+T^{-1}\big(U(Z)+ab^{-1}\big) \rd T \\
&=P(Z)^{-1}\big(T^{1/b}\rd X+T^{-1}b^{-1}Z\, \rd T\big),  
\end{align*} 
using (\ref{pred}) and (\ref{upaform}), hence 
\beq \label{zhod}
\rd z =\frac{1}{P(Z)}\, \rd Z.  
\eeq 
This defines a hodograph transformation between the solutions of (\ref{third}) and (\ref{logPZ}), where the latter can be rewritten as 
\beq\label{piiilike}
\frac{\rd^2 P}{\rd Z^2}=\frac{1}{P}\left(\frac{\rd P}{\rd Z}\right)^2-\frac{1}{Z}\left(\frac{\rd P}{\rd Z}\right)+\frac{1}{Z}\big( bP^b+a) -\frac{1}{P}. 
\eeq
Hence we arrive at an analogue of Theorem \ref{travhod} for these scaling similarity reductions. 

\begin{thm}\label{redhod} 
The scaling similarity solutions of the $b$-family equation (\ref{meq}) for $b\neq 0$, which satisfy the equation (\ref{third}), 
are given in parametric form by $U=U(Z)$, $z=z(Z)$, where $U$ is given by (\ref{upaform}) in terms of 
the solution $P(Z)$ of the non-autonomous second order ODE (\ref{piiilike}),  and $z$ is determined from 
\beq\label{zform}
z(Z)=\log\left(\frac{\psi_+(Z)}{\psi_-(Z)}\right)+\mathrm{const}, 
\eeq
where $\psi_{\pm}$ are two independent solutions of the same Schr\"odinger equation, 
\beq \label{schrobar}
\left(
\frac{\rd^2}{\rd Z^2}+\bar{V}(Z) 
\right) \psi_\pm =0 
\eeq 
having Wronskian $W(\psi_+,\psi_-)=1$, subject to the requirement that 
$P(Z)=\psi_+(Z)\psi_-(Z)$, 
with the potential $\bar{V}(Z) $ being given in terms of $P=P(Z)$ by 
\beq\label{Vbar}
\bar{V}=-\frac{1}{4P^2}\left(\Big(\frac{\rd P}{\rd Z}\Big)^2-1\right) 
+ \frac{1}{2ZP}\left( 
\frac{\rd P}{\rd Z}-bP^b-a 
\right) 
.
\eeq  
\end{thm}

\begin{prf}
All of the preceding statements follow 
from direct application of the hodograph transformation (\ref{zhod}), as described previously, apart from the form of the 
Schr\"odinger equation
(\ref{schrobar}). 
The expression for the potential is obtained by replacing all of the $X$ derivatives in (\ref{ep}) by $Z$ derivatives, to find 
that under the reduction we have $V(X,T)=T^{2/b}\bar{V}(Z)$, where $\bar{V}(Z)$ is given in terms 
of $P(Z)$ by the right-hand side of (\ref{Veq}), and then the equation (\ref{piiilike}) can be used to eliminate 
the second derivative term, to yield the formula (\ref{Vbar}). 
\end{prf}

The form of the ODE  (\ref{piiilike})
is very similar to the third Painlev\'e equation (\ref{piii}). However, by directly applying 
Kowalewski-Painlev\'e analysis to the equation in the form (\ref{logPZ}), which is very similar to the corresponding analysis of 
the PDE (\ref{pXT}) carried out in \cite{h}, we see that $b=2,3$ are the only cases 
that have the Painlev\'e property. Indeed, movable singular points 
in (\ref{piiilike}) are obtained from leading order behaviour of the form 
$$ 
P\sim \ka (Z-Z_0)^\mu, 
$$ 
and for generic $b$ the only possible balances have the leading exponent 
$$ 
\mu =\frac{2}{1-b}\quad \mathrm{or} \quad \mu=1.
$$ 
If $b\not\in\Z$ then the $P^b$ term in the equation creates non-integer exponents in local 
series expansions with (at least one of) these leading order behaviours, 
and then the requirement that the leading exponent $2/(1-b)\in\Z$ implies that 
$b=2$ or $3$ are the only possibilities.

As we shall see, these two special cases both correspond to particular instances of (\ref{piii}). 
The non-autonomous Hamiltonian formulation of Painlev\'e equations 
was initially developed by Okamoto \cite{o}, but in \cite{hach} it was noted that a different 
type of Hamiltonian is required to cover the $b=2$ case of (\ref{piiilike}). Interestingly, the same 
sort of Hamiltonian formulation extends to all values of $b\neq 1$: if we take 
the Hamiltonian to be 
\beq\label{hamna}
h=rZ^{-1}P^2\pi^2 +\big( s+(1-as)Z^{-1}P\big)\, \pi -\frac{b}{2r(b-1)}P^{b-1}, 
\,\,
r\neq 0 \,\, \mathrm{arbitrary}, \,\, s=\pm 1,
\eeq
 with $\pi$ being the conjugate momentum to $P$, then 
Hamilton's equations 
$$ 
\frac{\rd P}{\rd Z}=\frac{\partial h}{\partial \pi}, \qquad 
\frac{\rd \pi}{\rd Z}=-\frac{\partial h}{\partial P}
$$ 
are equivalent to (\ref{piiilike}). (For the case $b=1$, the final term in (\ref{hamna}) 
should be replaced with $\log P$.) Fixing the scale so that  $r=1$ when $b=2$ 
corresponds to the choice made in \cite{hach}, which gives 
$h(Z)=\tfrac{\rd}{\rd Z}\log \tau(Z)$, where the tau function $\tau$ has simple 
zeros at movable poles/zeros of $P(Z)$.  

If we fix $z_0=0$ then for the ramp solution (\ref{Uramp}) we have $P(z)=(b+1)^{-1/b} z^{1/b}$, 
and applying the hodograph transformation 
 (\ref{zhod}) in reverse we find that 
\beq\label{alg} 
P(Z) =b^{-1/(b+1)} Z^{1/(b+1)}
\eeq 
is a solution of (\ref{piiilike}) when $a=0$, for any $b\neq -1$.

\begin{exa} {\bf The Camassa-Holm equation:}
When $b=2$, equation (\ref{piiilike}) is 
\beq\label{piiib2} 
\frac{\rd^2 P}{\rd Z^2}=\frac{1}{P}\left(\frac{\rd P}{\rd Z}\right)^2-\frac{1}{Z}\left(\frac{\rd P}{\rd Z}\right)+\frac{1}{Z}\big( 2P^2+a) -\frac{1}{P}, 
\eeq 
which 
is precisely the $\gam=0$ case 
of the third Painlev\'e equation; this reduction was first  obtained in \cite{hach}. We can identify the 
variables and parameters in (\ref{piii}) as follows: 
\beq\label{b2piii}
w=P, \quad \ze = Z, \quad \al=2, \quad \be =a, \quad \gam=0, \quad \delta = -1. 
\eeq   

It is known that the equation (\ref{piiib2}) admits a B\"acklund transformation:  for any solution $P=P(Z)$, the quantities 
\beq\label{bt}
P_\pm = \frac{Z\big(\pm P_Z+1\big)}{2P^2} + \frac{\big(\mp 1-a\big)}{2P} 
\eeq 
satisfy the same ODE but with the parameter replacement $a\to a\pm 2$. As pointed out in \cite{hach}, 
this B\"acklund transformation can be derived from the Darboux-Crum transformation for the Schr\"odinger equation 
(\ref{schrobar}). Moreover, if we take the Hamiltonian (\ref{hamna}) for $b=2$ with $r=1$ and the two possible choices of sign for $s$, then we 
find 
$$ 
h_\pm=Z^{-1}P^2P_\pm^2+(a\pm 1)Z^{-1}PP_\pm-P_\pm -P, 
$$ 
or in other words $\pi =\pm P_\pm$ is the conjugate momentum for each of the respective choices of sign.   

Adding the two equations (\ref{bt}) implies that 
$$ 
P_++P_-=\frac{Z}{P^2}-\frac{a}{P}, 
$$ 
but then using (\ref{piiib2}) this gives 
\beq\label{logder}
P_++P_-=2P-\frac{\rd}{\rd Z}\Big(Z (\log P)_Z\Big).
\eeq 
Now if we introduce a tau function $\si=\si(Z)$ such that 
$$ 
P=-\frac{\rd}{\rd Z}\Big(Z (\log \si)_Z\Big), 
$$ and similarly introduce tau functions $\si_\pm$ such that analogous relations hold for $P_\pm$, then the equation 
(\ref{logder}) implies that 
$$\frac{\rd}{\rd Z}\left(Z \Big(\log \Big[\frac{\si_+\si_-}{\si^2 P}\Big]\Big)_Z\right)=0, 
$$
which integrates twice to yield 
\beq\label{CD}
\frac{\si_+\si_-}{\si^2 P}=CZ^D,
\eeq
for some constants $C,D$.

If we express $P$ in terms of the tau function $\si$, then (\ref{CD}) becomes a bilinear equation of Toda type, that is 
$$ 
\si_+\si_-+CZ^D\left(\tfrac{Z}{2}\,\mathrm{D}_Z^2\, \si\cdot \si +\si \, \si_Z\right)=0, 
$$
where $\mathrm{D}_Z$ denotes the Hirota derivative. Thus  $\si_-,\si,\si_+$ should be viewed as adjacent tau functions at points $a-2,a,a+2$ on a lattice where each point is distance 2 away from the next.   

As a particular example of a sequence of solutions generated in this way, note that we can take $P=P_0(Z)=(Z/2)^{1/3}$ as a seed solution when $a=0$, corresponding 
to the ramp solution of the Camassa-Holm equation, and then applying the  B\"acklund transformation (\ref{bt}) both forwards and backwards produces a 
sequence of algebraic solutions $P_{2n}(Z)$ at parameter values $a=2n$ for $n\in\Z$, which are rational functions of $Z^{1/3}$ (see \cite{bcm} for a table 
with some of these solutions). In that case we find an associated  normalized sequence of tau functions $\si_{2n}$, such that $P_{2n}=-\tfrac{\rd}{\rd Z}\big(Z(\log\si_{2n})_Z\big)$, 
the corresponding potential $\bar{V}=\bar{V}_{2n}$ defined by (\ref{Vbar}) with $b=2$ is given by $\bar{V}_{2n}=2\big(\log\si_{2n}\big)_{ZZ}$, and the ratio 
$\si_{2n-2}\si_{2n+2}/(\si_{2n}^2 P_{2n})$ in (\ref{CD}) is equal to 1 
for $n$ even and  3 
for $n$ odd; some of these are listed in Table \ref{algch} below.
Of course, it is natural to set $Z=2\ze^3$ and rewrite everything in terms of polynomials in $\ze$. It appears that these solutions are not completely understood: for instance, 
apparently it is not known if these polynomials in $\ze$ can be written in terms of Wronskians of suitable Schur polynomials (see \cite{clarkson} and references for more details). 

\begin{table}[h!]
  \begin{center}
    \caption{Algebraic solutions  and tau functions for  (\ref{piiib2}) 
in terms of $\ze=(Z/2)^{1/3}$.}
    \label{algch}
\scalebox{0.9}{
    \begin{tabular}{ | r|| c| c| c| c |} %
\hline
      $a=2n$ & 0 & $\pm 2$& $\pm 4$ & $\pm 6$  \\
\hline 
&&&&\\
$P_{2n}$  & $\ze$ & $\frac{3\ze^2\mp 1}{3\ze}$         &   $\frac{\ze(9\ze^4\mp 12\ze^2 +5)}{(3\ze^2\mp 1)^2}$              
&      
$\frac{243\ze^{10}\mp 891\ze^8+1350\ze^6\mp 990\ze^4+315\ze^2\mp35}{3\ze(9\ze^4\mp 12\ze^2 +5)^2}$         
 \\ 
&&&&\\
\hline 
&&&&\\
$\si_{2n}$ & $\ze^{-\tfrac{5}{24}}e^{-\tfrac{9}{8}\ze^4}$
&  
$\ze^{\tfrac{7}{24}}e^{-\tfrac{9}{8}\ze^4\pm\tfrac{3}{2}\ze^2}$
&  
$\ze^{-\tfrac{5}{24}}\left(3\ze^2\mp 1\right)e^{-\tfrac{9}{8}\ze^4\pm 3\ze^2}$
& 
$\ze^{\tfrac{7}{24}}\left(9\ze^4\mp 12\ze^2 +5\right)e^{-\tfrac{9}{8}\ze^4\pm \tfrac{9}{2}\ze^2}$
\\ 
&&&&\\
\hline 
    \end{tabular}
}
  \end{center}
\end{table}

Note that, as is apparent from the above table, as $|Z|\to\infty$ all of these algebraic solutions are asymptotic to the solution $P_0=(Z/2)^{1/3}$, corresponding to the ramp profile of the 
Camassa-Holm equation. 
\end{exa}

\begin{exa}  {\bf The Degasperis-Procesi equation:} 
When $b=3$, equation (\ref{piiilike}) becomes 
\beq\label{piiib3} 
 \frac{\rd^2 P}{\rd Z^2}=\frac{1}{P}\left(\frac{\rd P}{\rd Z}\right)^2-\frac{1}{Z}\left(\frac{\rd P}{\rd Z}\right)+\frac{1}{Z}\big( 3P^3+a) -\frac{1}{P}, 
\eeq
which also corresponds  to an instance 
of  the third Painlev\'e equation, namely the case $\al=0$, after making a slight change of dependent and independent variables  (this was briefly mentioned in \cite{dhh1}, but never 
elaborated on). 
We can identify the 
variables and parameters in (\ref{piii}) as follows: 
\beq\label{b3piii}
w=\left(\frac{Z}{3}\right)^{-1/4} P, \quad \ze = 4 \left(\frac{Z}{3}\right)^{3/4}, \quad \al=0, \quad \be =\tfrac{4}{3}a, \quad \gam=1, \quad \delta = -1. 
\eeq   

It is well known that, in the generic case $\gamma\delta\neq 0$, the equation (\ref{piii}) can be rescaled so that 
it depends on only two essential parameters, which are associated with the root space $B_2$, and the 
corresponding affine Weyl group acts birationally on the parameter space and the dependent/independent variables 
via B\"acklund transformations. 
Here we have chosen the normalization $\gam=-\delta=1$ as in \cite{bcm} (but see \cite{fw} for a 
different choice). In that case, given any seed solution with $\al=0$ and $\be$ arbitrary, we can use 
a composition of the B\"acklund transformation 
\beq\label{w1}
w^{(1)} =\frac{1}{w}-\frac{\al+\be+2}{\ze (w'+w^2+1)+(1+\al)w},  
 \eeq 
as in \cite{bcm}, with the prime denoting $\tfrac{\rd}{\rd \ze}$, which 
sends $\al\to\al+2$, $\be\to\be +2$, together with the transformation 
\beq\label{w2}
w^{(2)} =-\frac{1}{w}-\frac{\al-\be-2}{\ze (w'-w^2+1)+(1-\al)w},  
 \eeq 
which 
sends $\al\to\al-2$, $\be\to\be +2$, so that the overall 
effect is to send $\al\to\al$, $\be \to\be +4$ (and there are 
corresponding inverse transformations which can be combined to 
yield $\al\to\al$, $\be \to\be -4$); equivalently, one can use the composition of the two 
Schlesinger transformations $T_1,T_2$ for Painlev\'e III, as described 
e.g.\ in \cite{fw} (with a different choice of 
scaling for the parameters), which has the same overall effect: the main point is that one can keep the value 
$\al=0$ fixed, and just shift $\be$ up or down. In terms of the original ODE (\ref{piiib3}) obtained by reduction 
from Degasperis-Procesi, the effect is to shift the parameter $a\to a\pm 3$. 

For the case of (\ref{piiib3}), it turns out that there are various interesting choices of seed solution that can be used 
to generate explicit solutions for particular values of the parameter $a$. The simplest choice is the one corresponding to 
the  ramp solution, namely 
$P=P_0(Z)=(Z/3)^{1/4}$ for $a=0$. With the choice of normalization as in (\ref{b3piii}), this gives the constant seed solution 
$w=1$ for Painlev\'e III with parameters $\al=\be=0$ and $\gam=-\delta=1$, and the action of 
B\"acklund transformations on this solution generates solutions that are rational in $\ze$, which can be 
expressed in terms of so-called Umemura polynomials (see \cite{clarkson} and references for full details). If we 
apply the composition of (\ref{w1}) and (\ref{w2}), or the 
composition of their inverses, in order  to maintain the requirement that $\al=0$, then we get a 
particular sequence of these rational solutions for parameter values $\be=4n$, $n\in\Z$, and under the 
change of variables (\ref{b3piii}) this produces a sequence of similarity solutions 
for the Degasperis-Procesi equation which are given by functions $P_{3n}(Z)$ that 
are rational in $Z^{1/4}$, satisfying (\ref{piiib3}) at parameter values $a=3n$ (see Table \ref{algdp} below). 
Similarly to the case $b=2$, as $|Z|\to\infty$ all of these algebraic solutions are asymptotic to $P_0=(Z/3)^{1/4}$, 
corresponding to the ramp profile for the Degasperis-Procesi equation.

\begin{table}[h!]
  \begin{center}
    \caption{Algebraic solutions  for  (\ref{piiib3}) 
in terms of $\ze=4(Z/3)^{3/4}$.}
    \label{algdp}
\scalebox{1.0}{
    \begin{tabular}{ | c| c| } %
\hline
      $a=3n$ & $P_{3n}$    \\
\hline 
\hline
&\\
0 
& 
$\left(\frac{\ze}{4}\right)^\frac{1}{3}$
\\
& \\
\hline 
 & \\
3 & 
 $\left(\frac{\ze}{4}\right)^\frac{1}{3}\left(\frac{2\ze-3}{2\ze-1}\right)$             
\\
& \\
\hline 
& \\ 
6 &   
 $\left(\frac{\ze}{4}\right)^\frac{1}{3}\frac{(2\ze-3)(8\ze^3-60\ze^2+150\ze-105)}
{(2\ze-5)(8\ze^3-36\ze^2+54\ze-15)}$ 
\\
& \\ 
\hline 
& \\ 
9& 
$\left(\frac{\ze}{4}\right)^\frac{1}{3}\frac{(8\ze^3-60\ze^2+150\ze-105)
(64\ze^6-1344\ze^5+11760\ze^4-53760\ze^3+132300\ze^2-162540\ze+72765)}
{(8\ze^3-84\ze^2+294\ze-315)
(64\ze^6-960\ze^5+6000\ze^4-192000\ze^3+31500\ze^2-23940\ze+4725)}$       
 \\
&\\
\hline 
    \end{tabular}
}
  \end{center}
\end{table}

Painlev\'e III also admits one-parameter families of classical solutions in terms of Bessel functions. 
With the choice of scaling in \cite{fw}, the parameters in (\ref{piii}) are given by 
\beq\label{roots} 
\al = -4v_2, \quad \be = 4(v_1+1), \quad \gam=-\delta =4, 
\eeq 
where the pair $(v_1,v_2)$ is associated with the $B_2$ root space. The classical solutions 
are obtained by starting from the line $v_1+v_2=0$ in parameter space. Along this line, there are special solutions such that the function 
$w$ satisfies a Riccati  equation, and linearizing the latter shows that such $w$ are given in terms of 
the logarithmic derivative
 of the solution of a linear equation equivalent to Bessel's equation with parameter $v_1$; so for $v_1\not\in\Z$, this 
can be written using a linear 
combination of the modified Bessel 
functions $I_{\pm v_1}$ with argument proportional to $\ze$: the reader is referred to Proposition 
4.3 in \cite{fw} for the precise details. For our purposes, the main point is to see how this relates to particular solutions of 
(\ref{piiib3}). Upon comparing the choice of scale in (\ref{roots}) with (\ref{b3piii}), we see that the 
requirement $\al=0$ together with $v_1+v_2=0$ fixes $v_1=v_2=0$, while in general the parameter $a$ is related to $v_1$ 
by $a=\tfrac{3}{2}(v_1+1)$, so we obtain a Riccati equation for $P$ at the parameter value $a=\tfrac{3}{2}$,  
with a one-parameter family of solutions in terms of a combination of the Bessel functions $J_{0}$ and $Y_0$. 
Then by applying the composition of the two transformations (\ref{w1}) and (\ref{w2}), or their inverses, 
starting from a seed solution of this kind with $a=\tfrac{3}{2}$, we obtain a 
sequence of related solutions of (\ref{piiib3}) at parameter values $a=3n+\tfrac{3}{2}$ for $n\in\Z$. 
Note that in fact it is sufficient to just derive the solutions for non-negative integers $n$, since the ODE for $P$ 
has the discrete symmetry $P\to -P$, $a\to -a$; so  for negative $n$ the solutions are found immediately by applying this symmetry (and the same consideration applies to the algebraic solutions in Table \ref{algdp}). 
\end{exa}

\begin{remark}
The equation (\ref{piiilike}) with $b=-1$, that is 
$$ 
\frac{\rd^2 P}{\rd Z^2}=\frac{1}{P}\left(\frac{\rd P}{\rd Z}\right)^2-\frac{1}{Z}\left(\frac{\rd P}{\rd Z}\right)+\frac{a}{Z} -\left(1+\frac{1}{Z}\right)\frac{1}{P}, 
$$ 
is extremely close to the special case $\al=\gam=0$ of Painlev\'e III, which is one of the degenerate cases where 
(\ref{piii}) can be reduced 
to a quadrature and the general solution given in terms of elementary functions (see \cite{bcm}, for 
instance). However, the presence of the additional final  term $1/(ZP)$ above means that reduction to a quadrature is 
no longer possible, and perhaps the best that can be done is to produce asymptotic series solutions for this $b=-1$ 
equation in the limit $|Z|\to \infty$. 
\end{remark} 

\section{Conclusions} 

\setcounter{equation}{0}

We are planning at least one article in the near future, in which we propose to describe the details of analogous scaling similarity reductions for other peakon equations. 
In particular, in \cite{lucy} we have obtained related results for two integrable 
peakon equations with cubic nonlinearity, namely the equation 
\beq\label{forq} 
m_t+\Big(m(u^2-u_x^2)\Big)_x=0, \qquad m=u-u_{xx}, 
\eeq 
which was derived in \cite{fokas} and \cite{or}, and considered more recently in \cite{qiao},  
as well as Novikov's equation 
\beq\label{vn}
m_t+u^2m_x+ 3uu_xm=0, \qquad m=u-u_{xx},
\eeq 
which was 
obtained from a classification of such equations admitting infinitely many local symmetries in $m$ \cite{novikov}. 
It turns out that both of these equations admit similarity reductions that are connected via a hodograph transformation 
to certain equations of Painlev\'e type: for the reductions of (\ref{forq}), an equation of second order and second degree arises, 
while for (\ref{vn}) one finds a 
special case of the Painlev\'e V equation. Moreover, it happens that both of these reductions 
can be solved in terms of solutions of Painlev\'e III, so that the reduction of 
(\ref{forq}) is related to (\ref{piiib2}), while scaling similarity solutions of (\ref{vn})  are precisely the special cases of Painlev\'e 
V transcendents that are related to  Painlev\'e III in the form (\ref{piiib3}). These 
connections are   
not entirely surprising in the light of the fact that, in a certain sense, (\ref{forq}) can be considered as a modified 
Camassa-Holm equation, while (\ref{vn}) can be viewed as a modified version of the Degasperis-Procesi equation. 
However, the relevant connections with the cases $b=2, 3$ of (\ref{bfam}) are far from being straightforward, since 
reciprocal transformations are involved. 

We also hope to obtain a more explicit description of the algebraic solutions in Table \ref{algch}, using the Crum 
transformation for the corresponding  Schr\"odinger equation (\ref{schrobar}), since it appears that a
determinantal formula for the associated sequence of special polynomials in $\zeta$ is currently lacking.

\noindent \textbf{Acknowledgments:} LEB was supported by a PhD studentship from SMSAS, Kent. The research of ANWH was supported by 
Fellowship EP/M004333/1 from the
Engineering \& Physical Sciences Research Council, UK, and is currently funded by 
grant 
 IEC\textbackslash R3\textbackslash 193024 from the Royal Society. 
Conflict of Interest: The authors declare that they have no
conflicts of interest.


\end{document}